\documentclass[a4paper,fleqn,usenatbib]{mnras}

\usepackage[T1]{fontenc}
\usepackage{ae,aecompl}

\usepackage{graphicx}	

\usepackage{savesym}
\usepackage{amsmath}
\savesymbol{iint}
\savesymbol{iiint}
\usepackage{txfonts}
\restoresymbol{TXF}{iint}
\restoresymbol{TXF}{iiint}

\title[Microvariability in type 2 QSOs]{Detecting microvariability in type 2 quasars using\\
 enhanced F-test}

\author[J. Polednikova et al.]{
J. Polednikova,$^{1,2}$
A. Ederoclite,$^{3}$
J.A. de Diego,$^{4,5}$
J. Cepa,$^{1,2}$
\newauthor J.I. Gonz\'alez-Serrano,$^{6}$
A. Bongiovanni,$^{1,2}$
I. Oteo,$^{7,8}$
A.M. P\'erez Garc\'ia,$^{1,2,9}$ 
\newauthor R. P\'erez-Mart\'inez,$^{10, 11}$
I. Pintos-Castro,$^{1,2,12}$
M. Ram\'on-P\'erez,$^{1,2}$
\newauthor and M. S\'anchez-Portal,$^{11,13}$
\\
$^{1}$Instituto de Astrof\'isica de Canarias, C/Via Lactea s/n, La Laguna, 38205 Spain\\
$^{2}$Departamento de Astrof\'isica, Universidad de La Laguna, La Laguna, Spain\\
$^{3}$Centro de Estudios de F\'isica del Cosmos de Arag\'on, Teruel, Spain\\
$^{4}$Instituto de Astronom\'ia, Universidad Nacional Aut\'onoma de M\'exico, M\'exico D.F., M\'exico\\
$^{5}$Instituto de Astrof\'isia de Canarias - Universidad de La Laguna, CEI Canarias: Campus Atl\'antico Tricontinental, La Laguna, 38205 Spain\\
$^{6}$Instituto de F\'isica de Cantabria (CSIC-Universidad de Cantabria), Santander, Spain\\
$^{7}$Institute for Astronomy, University of Edinburgh, Royal Observatory, Blackford Hill, Edinburgh EH9 3HJ\\
$^{8}$European Southern Observatory, Karl-Schwarzschild-Str. 2, 85748 Garching, Germany\\
$^{9}$ASPID Association, Ap. correos 412, La Laguna, Spain\\
$^{10}$XMM/Newton Science Operations Centre, ESAC/ESA. Villanueva de la Ca\~nada, Madrid, Spain\\
$^{11}$Ingenier\'ia de Sistemas para la Defensa de Espa\~na (Isdefe), Madrid, Spain\\
$^{12}$Centro de Astrobiolog\'ia, INTA-CSIC, Villanueva de la Ca\~nada, Madrid, Spain\\
$^{13}$Herschel Science Centre, ESAC/ESA, Villanueva de la Cañada, Madrid, Spain
}

\date{Accepted XXX. Received YYY; in original form ZZZ}

\pubyear{2016}

\begin{document}
\label{firstpage}
\pagerange{\pageref{firstpage}--\pageref{lastpage}}
\maketitle

\begin{abstract}
Microvariability  (intra-night variability) is a low amplitude flux change at short time scales (i. e. hours). It has been detected in unobscured type 1 AGNs and blazars. However in type 2 AGNs, the detection is hampered by the low contrast between the presumably variable nucleus and the host galaxy. 
In this paper, we present a search for microvariability in a sample of four type 2 quasars as an astrostatistical problem. We are exploring the use of a newly introduced enhanced F-test, proposed by \citet{die14}.
The presented results show that out of four observed target, we are able to apply this statistical method to three of them. Evidence of microvariations is clear in the case of quasar J0802+2552 in all used filters ($g′$ ,$r′$ and $i′$) during both observing nights, and they are present in one of the nights of observations, J1258+5239 in one filter ($i'$), while for the J1316+4452, there is evidence for microvariability within our detection levels during one night and two filters ($r'$ and $i'$).
We demonstrate the feasibility of the enhanced F-test to detect microvariability in obscured type 2 quasars. At the end of this paper, we discuss possible causes of microvariability. One of the options is the misclassification of the targets. A likely scenario for explanation of the phenomenon involves optically thin gaps in a clumpy obscuring medium, in accordance with the present view of the circumnuclear medium. There is a possible interesting connection between the merging state of the targets and detection of microvariability.
\end{abstract}




\section{Introduction}

Active Galactic Nuclei (AGN hereafter) can be explained by the standard (unified) model, introduced by \citet{ant}. An AGN is described as a central supermassive black hole, surrounded by an accretion disc of an optically thick plasma emitting great amounts of energy in UV and X-ray wavelengths, and by an optically thin region of free fall onto the central black hole, where hard X-rays and gamma rays are emitted. On top of the accretion disc, there is a corona of energetic electrons, able to scatter the light via the inverse Compton effect. All of this is surrounded by an obscuring torus of dust and gas, located at a greater distance from the central singularity.

Different classes of AGNs are explained by the unified scenario as they are solely an orientation effect of the obscuring torus. Based on the unified model, the central engine of type 1 AGNs is directly observable. These include Seyfert 1 galaxies and type 1 quasars (in the cases of low and high luminosities, respectively). Type 2 sources include Seyfert 2 galaxies and, in the case of higher luminosities, type 2 quasars. Aside from these AGN types, we can also include blazars which are the sources observed face-on, which do not show dust and gas obscuration and are dominated by the non-thermal synchrotron emission from the jet.

The obscured sources are detected mostly via X-ray selection \citep{iwa}, from ultra luminous infrared galaxies \citep[ULIRGs;][]{tra}. Recently, \cite{rey08} established a new proxy for identifying type 2 quasars in the optical regime, replacing the need of bolometric luminosity which is accessible with difficulty in  type 2 objects due to the presence of the obscuring torus. Instead they proposed measurements of [OIII] luminosities, which is more easily accessible. This way, obscured AGNs with $L_{\rm{[OIII]}} \geq 10^{8.3}L_{\odot}$ are classified as type 2 quasars. 

The discovery of the variable behavior of quasars \citep{mas} followed along the discovery of quasars. The variable behavior is present also in the lower luminosity classes of AGNs \citep[Seyfert galaxies;][]{fic}. Variability is present in most, if not all, type 1 AGNs. The variable behavior in the lower luminosity classes of AGNs, along with the change in the characteristics of the optical spectrum of some of the Seyfert galaxies and LINERs from type 2 to type 1 \citep{khw, tom, coh, sbw, are} and vice versa, indicate that at least some of the type 2 AGNs are also variable, although these variations are attributed to changes in column density along the observer's line of sight rather than originating in the central engine. Recently there have also been reports of observing variability in the so-called unobscured type 2 quasars in the long term time-scale \citep{li}.

The amplitude of the variability depends on the time-scale of the phenomenon as well as on the observed wavelength. The amplitude is increasing with the increasing frequency and shortening of the wavelength. In the optical regime, variability is detected on  time-scales which can range from minutes to years \citep{kid, car, gok, jam95, ram}. 

We define microvariability as variations of flux on a time scale from minutes to hours, with amplitudes $\sim 0.05$ mag. The first microvariability detections were reported in \citet{mas} - variations of 15 minutes in 3C 48 (type 1 quasar). However due to small amplitude of the variations and the limitations imposed by the instruments, the results were considered unreliable until the development of the CCDs that allowed more precise photometry. Nevertheless microvariability studies mostly covered blazars \citep[e. g.][]{mil} or type 1 AGNs, with the emphasis on the differences between radio loud (RL) and radio quiet (RQ) sources. It was proposed that the microvariable behavior is common between the radio loud sources \citep{jam95,jam97}. According to \citet{gok03,sta,car07}, there is smaller fraction of microvariable sources among RQ targets. However, extensive studies by \citet{die98} and \citet{ram09} showed that there is no significant difference between the RQ and RL AGNs in terms of microvariability detection. The mentioned studies were dedicated to unobscured sources. As according to the causal arguments, we can assume that the microvariations are originating close to the central engine of the AGN, directly accessible in blazars and type 1 sources.  Following this assumption, microvariability helps us  gain insights about the innermost regions close to the central black hole. 

As the microvariations are attributed to the innermost regions of the AGN, their detection in type 2 sources presents an opportunity to probe regions otherwise inaccessible or very challenging to access. Nevertheless, the detection of microvariability is challenging on its own. Currently, the analysis of the photometric data for microvariability relies on a statistical approach. Various statistical tests were used in the past to get an estimation whether the source is variable or not, amongst the most notable ones, C-test \citep{jam97}, one-way analysis of variance \citep[ANOVA;][]{die98} and F-test \citet{how} and its modification \citep{jos, joc, pal, goy}. Nevertheless, not all of the mentioned test are suitable and powerful enough to detect microvariability, as discussed in \citet{die14}. 

In this paper, we are presenting the results from a campaign dedicated to searching microvariability in a sample of four type 2 quasars, using the enhanced F-test as proposed by \citet{die14}. The paper is organized as following: section 2 describes the sample and the data reduction, section 3 describes in the detail the enhanced F-test, section 4 contains discussion about the results obtained on every target and section 5 provides conclusions.

\section{Sample, observations and data analysis}

\subsection{Sample selection}

The sample was chosen from the catalog of type 2 quasars by \citet{rey08}. This catalog is based on the Sloan Digital Sky Survey Data Release 6 \citep{adc}. With 887 sources, this is the biggest collection of type 2 quasars to date. We have restricted the sources to be bright ($g' <18$ mag) and with redshift $z < 0.1$. By restraining the redshift range, we are avoiding differences which might arise due to a change of the rest frame of the targets. This effect might result in an unnecessary bias of monitoring different physical regions. The spread over the redshifts is 0.0367, which results in $\Delta\lambda \approx 144 $\AA. This difference is covered within one broad band filter.  

The restriction placed on the brightness is based on the aim to obtain  high quality data with signal to noise ratio $> 100$. There are 14 targets meeting our criteria. The final sample was chosen based on the visibility during the observing nights. The properties of the selected targets are reported in table \ref{Table1}.

\begin{table}
\caption{Overview of the observed sample. Galaxy J1316+4452b is neighbor of J1316+4452 which was observed in the same field as the targeted type 2 QSO J1316+4452. It is described as broad line galaxy in SDSS-III survey.
\label{Table1}}
\begin{center}
\begin{tabular}{c|cccc}
Target & RA & Dec & z& gmag\\
\hline
J0802+2552&08 02 52& +25 52 55&0.0811&15.93\\
J0843+3549&08 43 45& + 35 49 42&0.0539&15.28\\
J1258+5239&12 58 50& +52 39 12&0.0552&15.63\\
J1316+4452&13 16 39& +44 52 35&0.0906&15.98\\ 
J1316+4452b&13 16 36&+44 51 57& 0.0602 &16.17\\
\end{tabular}
\end{center}
\end{table}

\begin{figure*}
\centering
\includegraphics[width=17cm]{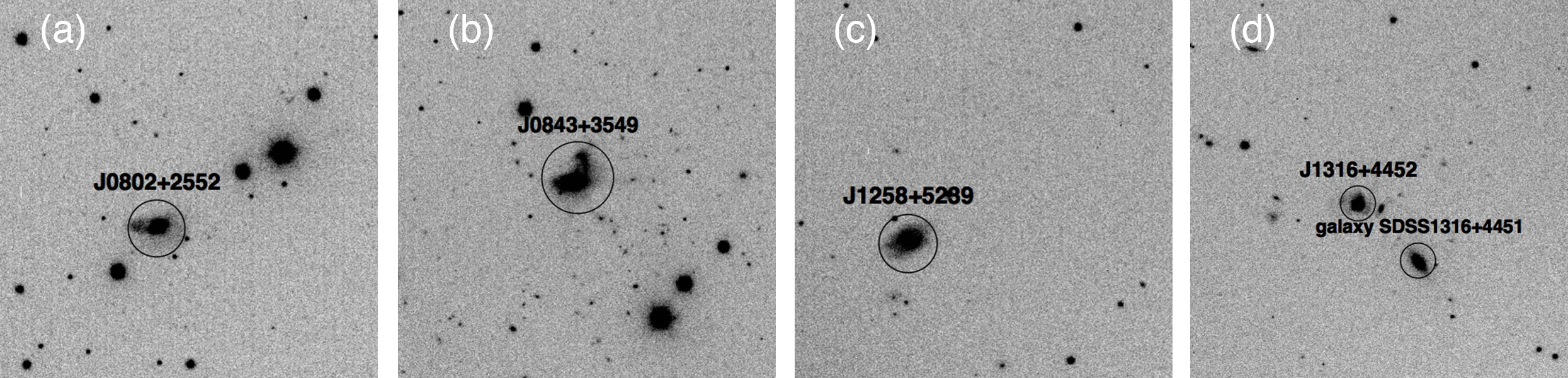}
\caption{Targets observed in our observing campaign are marked with a circle, mimicking the size of the aperture used for the photometry. Panels (a)~-~(b) show targets with disturbed morphologies. J1316+4452 shows accompanying galaxy of approximately same brightness as the source itself J1316+4452b. North is up in all the images, east on the right. \label{Fig1}.}
\end{figure*}

Although we have not initially focused on the morphology, it should be noticed that two of our targets have highly disturbed morphologies (Figure \ref{Fig1} Panels a and b), pointing towards recent merging events. 

\subsection{Observations}

We have observed our targets on February 21 and February 22, 2014 with the 2.5 meter Nordic Optical Telescope on La Palma, equipped with the ALFOSC camera. Our observing strategy partly followed works by \citet{die98} and \citet{ram}. During the first observing night, we were observing solely in the Sloan $g'$ filter. During the second night, we were interchanging Sloan $g'$, $r'$ and $i'$ filters.  We have observed with $2 \times 2$ binning. We have reached a SNR $> 100$ in 60 seconds exposures in all filters. The targets were chosen to be close to each other, so that we could easily switch between them. The targets were switched every five exposures. Such setting allowed us to monitor the pairs on a longer time scale without long overheads. Previously detected microvariability events in quasars were reported to be of the time scales of  $\approx$ 1.5 hours \citep{ram, die98}. Based on this assumption, we have increased the chance of detecting any microvariations by setting the time on target to four hours per night.

Our monitoring resulted in roughly 70 points on the light curve per target during the first night, and 30 point per target per night in every filter during the second observing night. 

\subsection{Data Reduction}

Bias subtraction and flat-fielding were performed using the standard tasks in IRAF\footnote{Image Reduction and Analysis Facility (\texttt{http://iraf.net}).}. Aperture photometry was performed using SExtractor 2.8.6 \citep{ber}\footnote{Source Extractor (\texttt{http://www.astromatic.net/software\
/sextractor}).}. To avoid any possible systematic errors arising from  inhomogeneities in the CCD detector, which might be irremovable with flat-fielding, we used a dithering pattern when obtaining the images. Therefore arose the  need for precise  astrometry on every image. For this purpose we used SCAMP 2.0.4 \citep{scamp}\footnote{Software for Calibrating AstroMetry and Photometry (\texttt{http://www.astromatic.net/software/scamp}).}, which directly uses output from SExtractor. SCAMP matches the known pointing and SExtractor catalogs of the images with chosen catalog (in our case USNO-B2.0 catalog). Resulting astrometric catalogs for every image were matched using TOPCAT \citep{topcat}\footnote{Tool for OPeration on Catalogues And Tables (\texttt{http://www.star.bris.ac.uk/~mbt/topcat/}).}. 

The ALFOSC $i'$ images are affected by fringing  which needed to be removed prior to photometry. The pattern proved to be time variable, therefore we grouped the $i'$ filter observations in sets of five, following the observational pattern. The groups of five exposures were taken within 10 minutes, therefore we can assume that over such time scale, the pattern remains stable. 
The removal was performed using procedures natively available in IRAF. The pattern proved to be  stable within the groups of five exposures and therefore was successfully removed from all the $i'$ filter images.

We have used aperture photometry, even though our targets have resolved host galaxies. At the moment, there is no software which would allow us to decompose the galaxies which have complex irregular morphologies. Any attempt in decomposing our galaxies would most likely result in the introduction of additional errors. We have included the whole target in the aperture, including the  non variable host galaxy. Including non variable part would more likely underestimate the variation than enhance it. Besides the circular apertures, SExtractor offers also the option of elliptical apertures (AUTO and PETRO). Nevertheless the control of the size of the ellipse and its ellipticity is limited. On top of that, while the elliptical apertures might be more appropriate for galaxies, they are completely unjustified for stars. In order to obtain photometry as homogeneous as possible, we have used the same aperture diameter for the field stars. Unfortunately, the lack of precise photometric tools to measure heavily distorted galaxies such as the host of J0843+3549 prevents a reliable analysis of the variability for this object. For the details, see Section 4.


\section{Analysis}

We have analyzed our data using the enhanced F-test introduced by \citet{die14}, applied to differential photometry light curves. Such test already proved successful in analysis of blazar light curves \citep{gaur2015}. The enhanced F-test makes use of not only one star in the field, such as in the case of 'normal' F-test, where only one non variable star is used to compare with the source, thus increasing the power of the test and ability to detect finer variations. We are testing the null hypothesis that the target is not variable. We fail to reject the null hypothesis when $p > 0.001$. If we reject the null hypothesis, we accept the alternative hypothesis that the target is variable. However there are several steps which need to be undertaken before the computation of the F statistics. We describe the procedure in the subsections below.

\subsection{Estimating the error}

SExtractor is a very convenient and fast tool for photometry, but the photometric error  provided by this software often seem underestimated. 
A similar issue was addressed by other authors \citep[e.g.][]{goy13}, who introduced multiplicative factor for errors in the APPHOT package in IRAF. To our knowledge, there is no study which would estimate possible multiplicative factor for SExtractor. There is no evidence that even with the availability of this factor, it would be universal, independent of characteristics of the telescope, instruments or observed bands. Therefore to achieve the highest possible accuracy, we need to estimate the photometric errors independently on SExtractor. 

In order to fully benefit from the enhanced F-test, we need to calibrate the brightness of stars in the field. To obtain the relation for the photometric error, depending on the brightness, we have constructed differential light curves for all the stars present in the field along the observed quasar. Out of those, we have chosen a constant comparison star for the differential photometry, based on the light curve obtained in the previous step. If the light curve did not show any significant trend or  features, the star was considered constant at the first approximation.  The comparison star is of roughly similar brightness as the studied target. We have computed the standard deviation of every light curve which serves as the estimated photometric error for a given star brightness. Stars with high errors are likely variable, hence we can exclude them from our calibration field. The quasars themselves were also excluded from the fitting as we assume they might be variable and would show up as outliers. Figures 2-4  show that the errors are following an exponential trend. As the precision decreases towards the faint end, we have decided to fit only a handful of comparatively bright stars, using the weighted exponential. The error we can reach for an infinitely bright source gives us the maximal precision we can reach in our field, which is $<0.01$ magnitudes in most cases. 

\begin{figure}
\resizebox{\hsize}{!}{\includegraphics{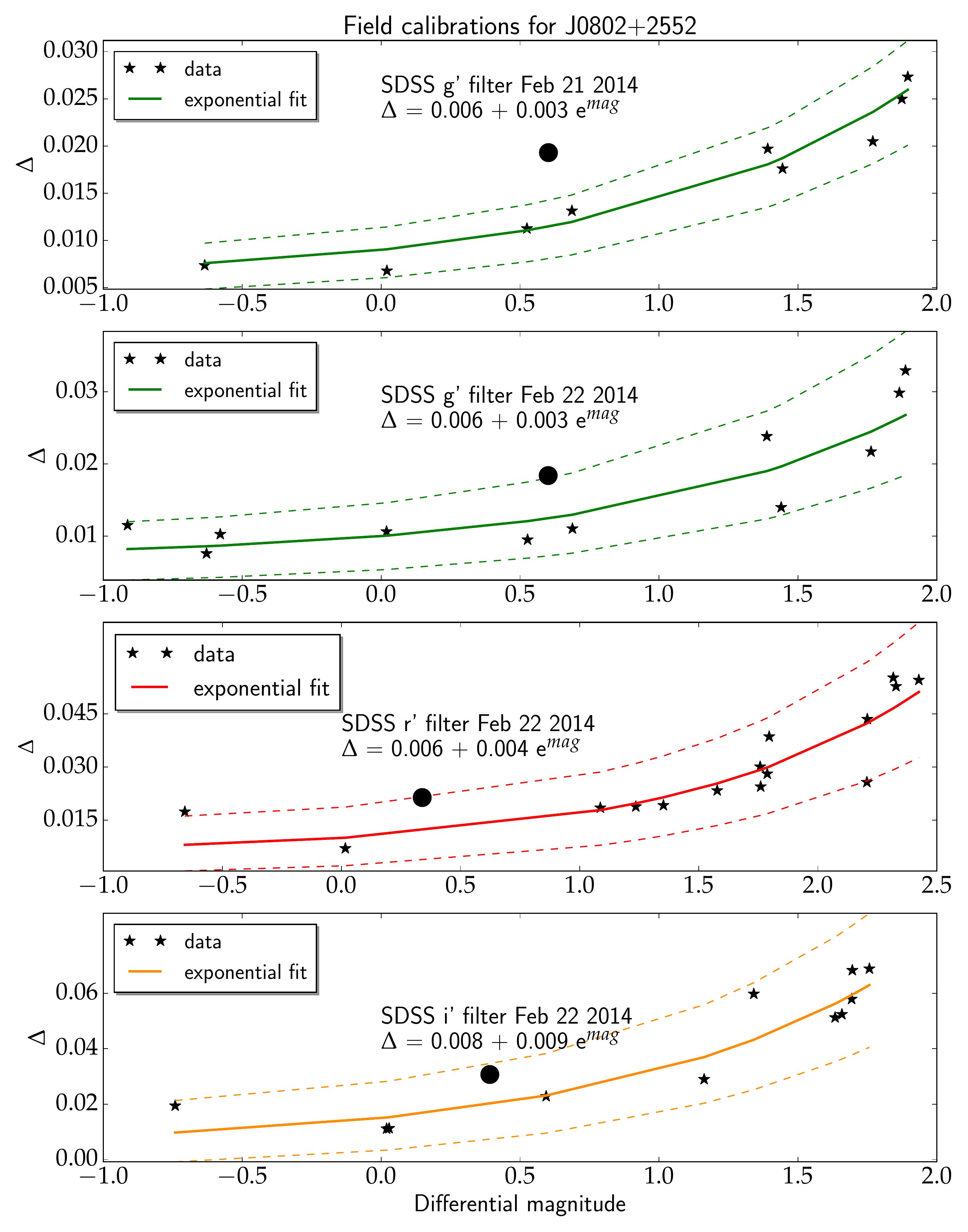}}
\caption{Calibration of the photometric error of the star field around J0802+2552. The $x$ axis marks the differential magnitudes of the field stars, $y$ axis is the photometric error, $\Delta$ predicted from the exponential fit. The dashed line marks 95\% confidence interval of the fit. The big black circle marks the position of the QSO (in all figures). First panel corresponds to February 21, 2014, when we observed solely in $g'$ filter, the three following panels correspond to $g'$,$r'$ and $i'$ filters observations on February 22, 2014. The graphs are color-coded, according to the filters ($g'$ green, $r'$ red and $i'$ orange; color version available only in the electronic journal) \label{Fig2}.}
\end{figure}

\begin{figure}
\resizebox{\hsize}{!}{\includegraphics{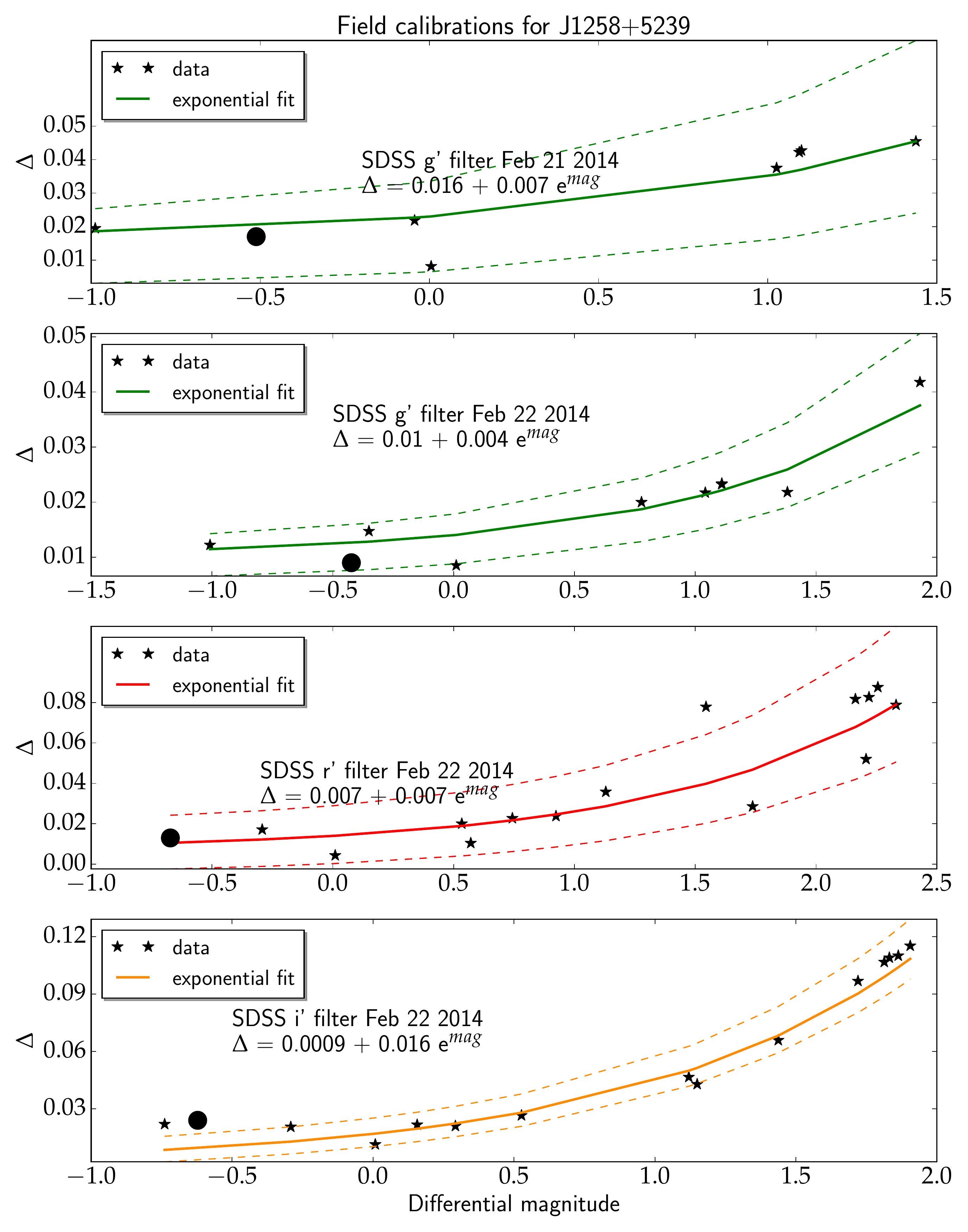}}
\caption{ Calibration of the photometric error of the star field around J1258+5239. The black circle marks the differential magnitude of the QSO with error $\Delta$. The axes and color coding is adapted from the previous figure. \label{Fig3}.}
\end{figure}

\begin{figure}
\resizebox{\hsize}{!}{\includegraphics{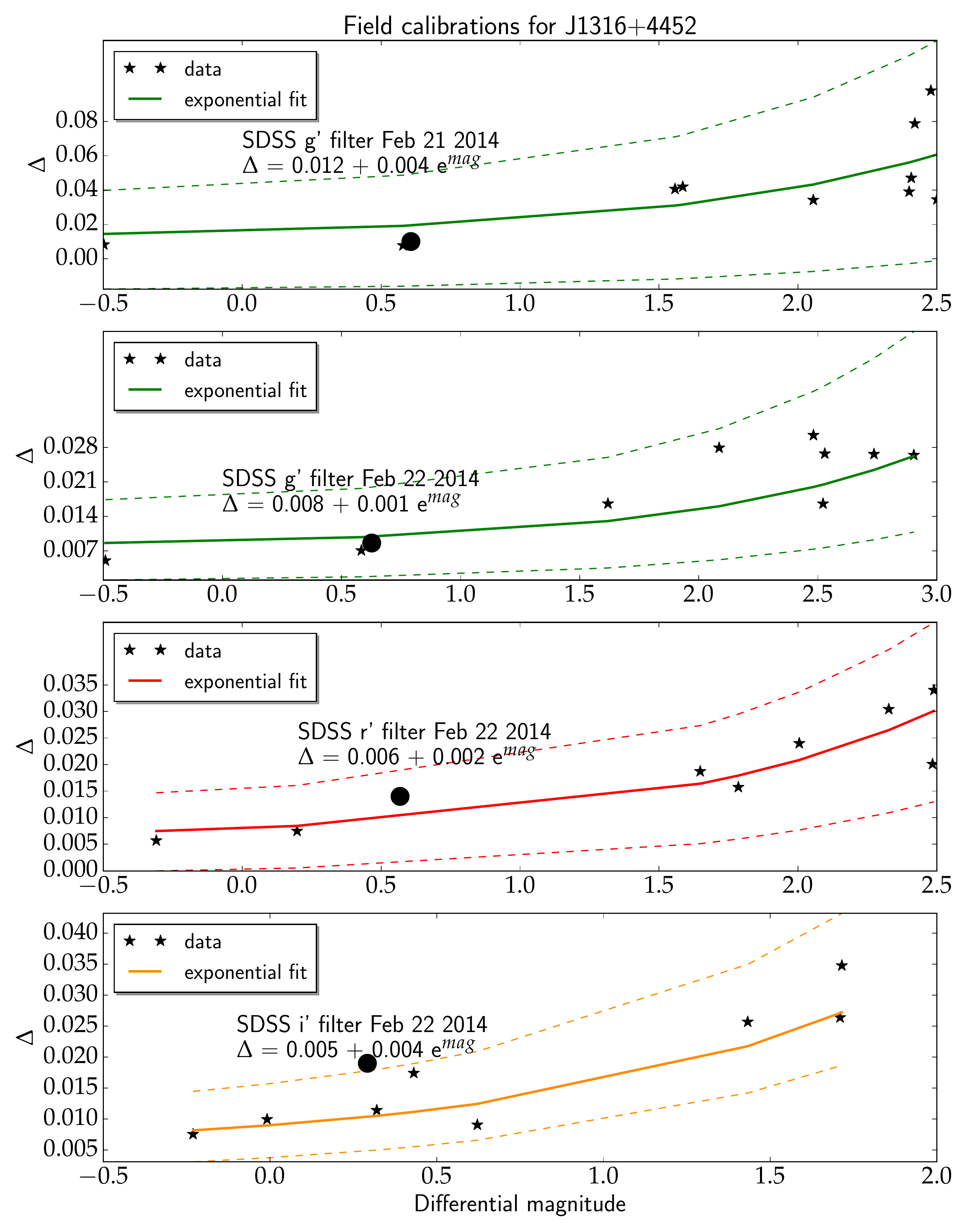}}
\caption{ Calibration of the photometric error of the star field around J1316+4452. The $x$ axis marks the differential magnitudes of the field stars, $y$ axis is the photometric error, $\Delta$ predicted from the exponential fit. The black circle marks the position of the QSO. Color-coding is following the same philosophy as in the previous figures. The field contains fewer stars bright enough to be fitted in comparison with the other targets, especially in $i'$ filter \label{Fig4}.}
\end{figure}

\subsection{Errors scaling}

In the ideal case, our comparison field stars would be of the same brightness as the targeted quasar. As it is not the case, we have to employ a scaling factor, $\omega_j$ to scale variance of the $j$-th star to the level of quasar $q$, where $\omega_j$ is determined based on the values obtained from the fits in Figures \ref{Fig2}, \ref{Fig3} and \ref{Fig4}. \citet{how} and \citet{jos} proposed different approaches to determine this scaling factor. As \citet{die14} remarks, it is not necessary to have the same number of observations of all comparison stars. Let's assume $k$ number of stars observed in the field and $N_j$ number of the observations of the $j-$th star. Observations are indexed over $i$. We will estimate the pooled (combined) variance. We are using the scaled magnitudes of the stars, but this does not affect the estimation of pooled variance. For $N_j$ stars, we can calculate the pooled variance as

\begin{equation}
\begin{aligned}
s_c^2 &= \frac{\sum^k_{j=1} (N_j - 1)s_j^2}{\sum^k_{j=1}(N_j - 1)}\\ 
&= \frac{(N1 - 1)s_1^2 + (N_2 -1)s_2^2 + \dots + (N_k - 1)s_k^2}{(\sum^k_{j=1} N_j) - k},
\end{aligned}
\end{equation}

\noindent where $s_j^2$ are the transformed light curve variance for the star $j$:

\begin{equation}
s_j^2 = \frac{\sum_{i=1}^{N_j}(x_{j,i} - \overline{x_j})^2}{N_j - 1}
\end{equation}

\noindent and the $(x_{j,i} - \bar{x_j})^2$ are the square deviations for the scaled magnitudes of the stars:

\begin{equation}
(x_{j,i} - \overline{x_j})^2 = \omega_j (m_{j,i}-\overline{m_j})^2 = s^2_{j,i}
\end{equation}

\noindent where $m$ denotes the magnitudes of the observed stars. $\overline{m_j}$ is the mean magnitude of the light curve of each star. 

Equations (2) and (3) yield:

\begin{equation} 
\sum_{i=1}^{N_j} s_{j,i} ^2 = (N_j -1) s_j^2,
\end{equation}

\noindent which can be rearranged as  

\begin{equation}
s_c^2 = \frac{1}{(\sum^k_{j=1}N_j)-k}\sum^k_{j=1}\sum^{N_j}_{i=1} s^2_{j,i}.
\end{equation}

 We have found that the empirical relation for the photometric errors, which for a non-variable object are distributed as the standard deviation of the light curve $s$, is in the form 

\begin{equation}
s_{star} = s_j= a + b\,e^{\overline{m}},
\end{equation}

\noindent where$s_{star}$ is the standard deviation of the given star light curve. In the case of non-variable quasar, the $s_{qso}$ could be also fitted by equation (6). If the QSO is variable, it appears as an outlier from the fit. Fitting the dependence with a variable quasar results in a poor fit. We assume that the distribution of the observed magnitudes of non-variable comparison stars, is nearly normal, hence 

\begin{equation}
\frac{m_{star,i} - \overline{m}_{star}}{s_{star}^2} \sim \frac{m_{qso,i} - \overline{m}_{qso}}{s_{qso}^2} \sim \mathcal{N}(0,1),
\end{equation}

 \noindent Taking into account equation (6), we can derive the relation for computing the transformed magnitude, which is  

\begin{equation}
m_{t,star,i} = (m_{qso,i} - \overline{m}_{qso})\frac{s_{star}^2}{s_{qso}^2} + \overline{m}_{star},
\end{equation}

\noindent where $s_{qso}$ is the standard deviation of the quasar based on the fits. where the index $t$ denotes the transformed magnitudes. Taking into account the transformed magnitudes, we can  rewrite equation (3) as:

\begin{equation}
(x_{j,i} - \overline{x_j})^2 = (m_{t,j,i}-\overline{m_{t,j}})^2 = s^2_{j,i}.
\end{equation}

\subsection{F-test}

The F-test compares the variances of two samples. In our case we are testing whether our sample  of  stars with transformed magnitudes has similar variance as the quasar. Therefore 

\begin{equation}
F=\frac{s^2_{qso}}{s^2_{c}}, 
\end{equation}

\noindent where the variance $s^2_{qso}$ is obtained directly from the photomery. As we have scaled all the observations of all suitable field stars, we can greatly benefit from the high number of degrees of freedom which enhances the power of the F-test. Assuming $k$ is the total number of the scaled down stars and $N_{qso}$ is the total number of observations of the quasar, we have $\nu_1 = N_{qso} - 1$ degrees of freedom in the numerator and $\nu_2 = \sum_{j = 1}^{k}(N_j-1)$ in the denominator, where $N_j$ is the number of observations of the $i$-th star. We are loosing two additional degrees of freedom due to fitting of two parameters in equation (6). As we are scaling the stars to the brightness of the quasar, we can see that it is not necessary to observe all the calibrating field stars on every image. This allows us to dither images. The only star which needs to be present in all the images is the comparison star we are subtracting from all the field stars in order to obtain differential magnitudes.
To determine variability, we have computed the F test as in equation (10). To determine the success of the detection, we have computed cutoff levels for two level of significance; $\alpha = 0.01$ and $\alpha = 0.001$. Cutoff levels values are different for each target as we had different number of images and different number of field stars in every case. We have used the R-code\footnote{\texttt{http://www.r-project.org}} \citep{Rcore} to determine the values.

\section{Results}

The overall results for the separate filters are contained in the tables \ref{Table2}, \ref{Table3} and \ref{Table4}. We provide overview of the numbers of degrees of freedom used for the enhanced F-test in Table \ref{Table5}.

\begin{table*}
\caption{Results from enhanced F-test on sample in $g'$ filter. Y marks variability at the 0.1\% ($\alpha = 0.001$) significance level by default. F$_{\alpha =0.01}$ and F$_{\alpha = 0.001}$ mark the cutoff values for claiming the variability, given the degrees of freedom used. Column marked 'F$_{star}$' provides F test for a check star from the same field, measured with the same method as the quasars. The variability result is marked in the var? column, Y is variability detected at $\alpha = 0.001$, P means that the source is probably variable (the variability is detected at $\alpha = 0.01$) and N signifies no variability detected.
\label{Table2}}
\begin{center}
\begin{tabular}{c|cccccc||cccccc}
\multicolumn{1}{c|}{} & \multicolumn{6}{c||}{February 21, 2014}&\multicolumn{6}{c}{February 22, 2014}\\
\hline
Target & F$_{star}$ & F$_{QSO}$ & F$_{\alpha =0.01}$ & F$_{\alpha = 0.001}$& var? &p-value& F$_{star}$ & F$_{QSO}$ & F$_{\alpha =0.01}$ & F$_{\alpha = 0.001}$&var? &p-value\\
\hline
J0802+2552 & 1.19 & 2.93 & 1.52 & 1.74& Y&1.54e-10& 1.41 &2.07&1.77&2.11&P&2.0e-3\\
J1258+5239 & 1.16& 1.02 & 1.44 & 1.62&N&0.43& 1.00&0.56&1.81&2.17&N&0.96\\
J1316+4452 & 0.16&0.31 & 1.44 & 1.61 &N& 0.31 &0.58&0.58 &1.81 & 2.17 & N &0.99\\
J1316+4452b& 0.16 &0.25 & 1.44 & 1.61 & N&0.99& 0.40&0.55 & 1.80 & 2.17 &N&0.99\\
\end{tabular}
\end{center}
\end{table*}

\begin{table*}
\caption{Results from the observations taken on February 22, 2014 in $r'$ filter. The columns follow the same arrangement as in the previous table
\label{Table3}}
\begin{center}
\begin{tabular}{c|cccccc}
Target & F$_{star}$&F$_{obs}$ & F$_{\alpha =0.01}$ & F$_{\alpha = 0.001}$& var?& p-value\\
\hline
J0802+2552&0.59 &2.45&1.76&2.10&Y&7.0e-05\\
J1258+5239& 0.89&1.30&1.78&2.12&N&0.14\\
J1316+4452& 0.77&2.36&1.83&2.21&Y&3.8e-4\\
J1316+4452b& 0.77&4.03&1.83&2.21&Y&6.3e-09\\

\end{tabular}
\end{center}
\end{table*}

\begin{table*}
\caption{Results from the observations taken on February 22, 2014 in $i'$ filter. The columns follow the same arrangement as in the previous tables for $g'$ and $i'$ filters
\label{Table4}.}
\begin{center}
\begin{tabular}{c|cccccc}
Target & F$_{star}$& F$_{obs}$ & F$_{\alpha =0.01}$ & F$_{\alpha = 0.001}$& var?&p-value\\
\hline
J0802+2552& 1.49 &3.04&1.78&2.12&Y&1.1e-06\\
J1258+5239& 0.64&4.96&1.76&2.10&Y&1.1e-11\\
J1316+4452& 1.01&2.87&1.85&2.25&Y&2.2e-05\\
J1316+4452b& 1.01&3.73&1.85&2.25&Y&1.1e-07\\
\end{tabular}
\end{center}
\end{table*}

\begin{table}
\caption{Numbers of degreed of freedom $df1$ and $df2$ used for the enhanced F-test. The number of degrees of freedom of the star ($df2$) does not correspond to the expected number based on multiplication of the exposures by the number of stars used. This is caused by a slight shift in the observations. It does not affect the results. \label{Table5}.}
\begin{center}
\begin{tabular}{c|c|c|cc}
Source & Date & Filter &$df1$ & $df2$\\
\hline
J0802+2552 & 2014/02/21 &$g'$& 59 & 504\\
& 2014/02/22 &$g'$& 29 & 313 \\
& 2014/02/22 & $r'$& 29 & 402\\
&2014/02/22& $i'$& 29 & 305\\
\hline
J1258+5239 & 2014/02/21 & $g'$ & 84 & 479\\
& 2014/02/22 & $g'$ & 29 & 202\\
& 2014/02/22 & $r'$& 29 & 293\\
& 2014/02/22 & $i'$& 29 & 351 \\ 
\hline
J1316+4452 & 2014/02/21 & $g'$&83 &553\\
& 2014/02/22 &$g'$ & 29 &202 \\
& 2014/02/22 & $r'$& 29 & 163 \\
& 2014/02/22 & $i'$ & 29 & 135 \\
\end{tabular}
\end{center}
\end{table}

\textbf{J0802+2552} was observed during the first half of the night in both cases.  As we were observing just before the last quarter, the Moon was not affecting our observations.
The data from the first observing night, February 21 2014, were calibrated using 9 stars from the field. The omitted stars were either too faint or too bright and hence saturated (e.g. north-eastern part of Figure \ref{Fig1}). Using them would introduce imprecisions. We have obtained 60 exposures of the target in $g'$ band. We have used apertures of 34.4 arcsec (90 px). Not all the calibration stars were available in all the images, therefore the total number of calibration stars from the set of 60 images is 522 instead of 540. All of these stars were used as a reference to detect variability. We have detected variability at a significance level 0.001 during this observing night.

During the second night of the run, we have observed in three filters ($g'$, $r'$ and $i'$). We have obtained only 30 exposures per filter, therefore reducing the number of reference star data points and number of degrees of freedom. Nevertheless we had better quality observations and we were able to use more field stars (11 in $g'$, 15 in $r'$ and 11 in $i'$), hence the reduction in the number of exposures in a given filter with respect to the first night was partially compensated by the increase in the number of comparison stars. However the test lost some power, which is a possible reason why the target does not appear variable at the significance level 0.001. The variability is detected at the 0.01 level of significance. It is detected as variable at the significance level 0.001 in the $r'$ and $i'$ which both have higher number of field stars and therefore higher number of degrees of freedom. The aperture used for photometry was the same as during the first observing night.

\begin{figure*}
\centering
\includegraphics[width=17cm]{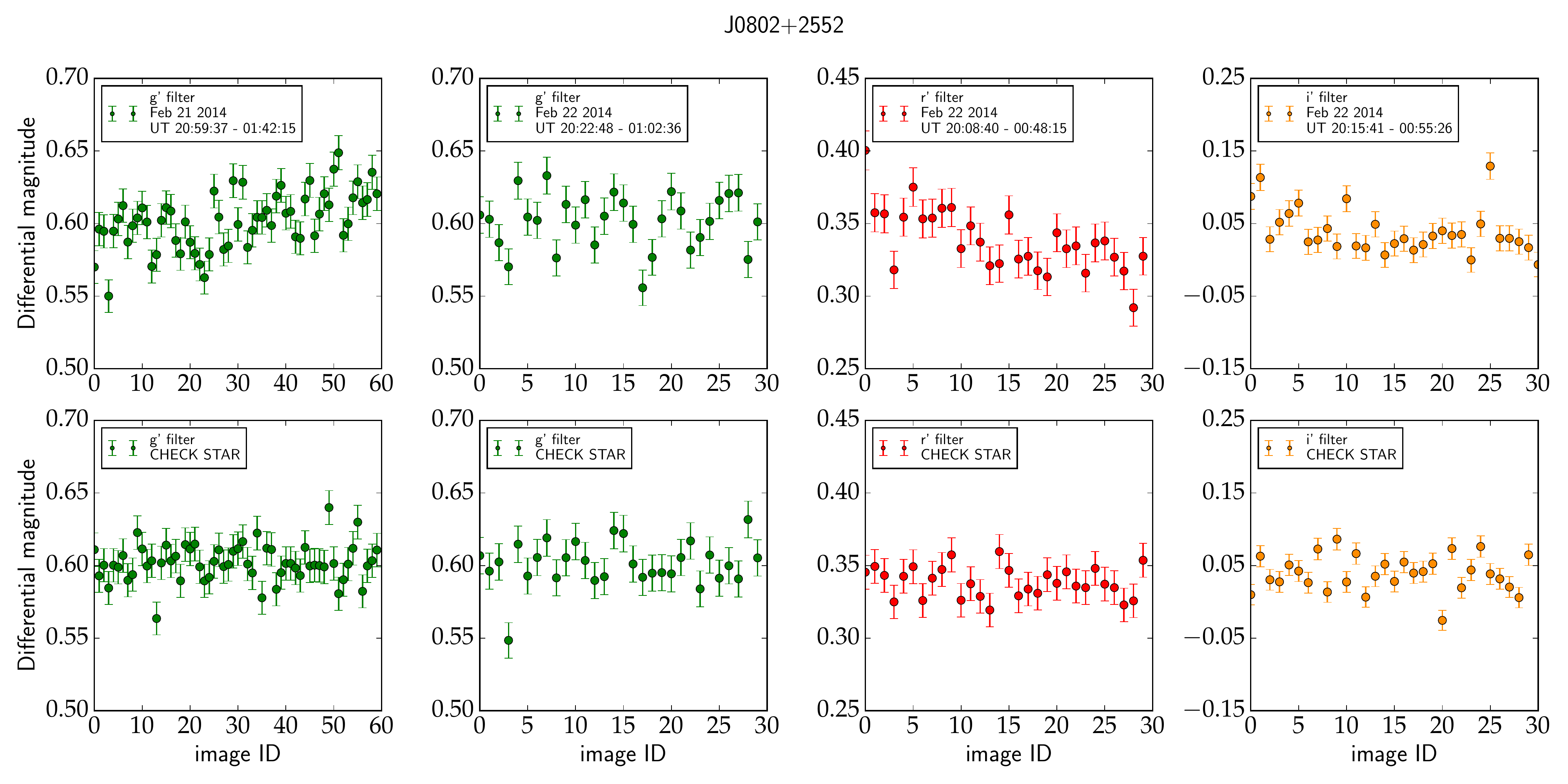}
\caption{Upper panels: Differential light curve for J0802+2552 during both nights in all available filters. The upper panel shows light curve of the QSO; the lower panel shows transformed differential light curve of the non-variable star used as check star in the test. There is a noticeable increase in the brightness in QSO $g'$ filter during the first night. During the second observing night, there is a decrease in brightness in both, $r'$ and $i'$ filters while $g'$ appears to be non variable. Color-coding is adopted from previous figures. The errors are computed in SExtractor \label{Fig5}.}
\end{figure*}

The light curves are displayed in figure \ref{Fig5}.  Overall we conclude positive detection of variability.

\textbf{J0843+3549} is a target with highly disturbed morphology. The irregular morphology presents a problem for the aperture photometry we are using in all the other cases. The target is quite extended, therefore to include it completely within the aperture is not feasible. Aperture diameter containing this target is approximately 45.6 arcsec (120 px). Such a big aperture is unsuitable for the stars. However the target, as shown in Figure \ref{Fig1}, seems to be a composite of two merging targets. We focus on the brighter (southern) target. SExtractor treats the whole source as two, the tail (in north-south direction) and the quasar itself,  therefore it seem sensible to use smaller aperture enclosing only one of the targets

The aperture photometry used by SExtractor counts the pixels above the detection threshold in the given aperture. The fainter tail is above the threshold level and as the seeing varied slightly over the first half of the night, we cannot control the propagation of the light from the tail into the target we are interested in. The varying propagation of the light would introduce higher variation of the target, which is not of physical origin. 

We have tried to constrain the threshold more strictly to avoid detection of the tail. The number of pixels above the threshold was increased to discard the tail. There are numerous caveats in this approach. We are severely lowering the number of stars which SExtractor detects. On the other hand, the bright stars are saturated. Therefore we fit only three stars which are roughly comparable in the brightness with the quasar. The aperture used in this case was 83 pixels (31.54 arcsec). To obtain the error estimate, we are fitting only three stars, which is the minimum number of stars to perform the F-test as pointed out by \citet{die14}. The asymptotic error is high in comparison with the other targets.

The resulting F statistic is very high (F $> 20$), providing very unlikely results. Comparing with the positive detections we have obtained in the other cases (F$\approx5$) it is highly possible that there is a fluctuation caused by light propagating into the aperture, which is not of a physical origin. Therefore we refrain from drawing conclusions about this target. Due to difficulties in performing accurate photometry, the results for the second observing night are also inconclusive.

\textbf{J1258+5239} was observed during the second part of the night, when the Moon had already risen. 

We have obtained 85 observations in $g'$ filter during the first night. The aperture photometry was performed with a diameter of 34.2 arcsec (90 px). We have used seven stars of the field. Taking into account the number of the exposures and the number of the field stars, considering that not all of them might be present on all the images due to the dithering pattern, we have obtained in total 535 scaled down star observations used for analysis of the variability. The statistical analysis results in negative detection of variability. The asymptote of the fit has a higher value compared with other days ($\approx 0.01$ mag for the infinitely bright source) which reduces our ability to detect microvariability. It borders the limit we set up for a sensible analysis. 

The seeing improved slightly during the second observing night, therefore we have reduced the size of the aperture to 28.5 arcsec (75 px).  We have obtained 30 images in all filters. Accounting for this change with respect to the first night, following the strategy for J0802+2552, we have used more field stars (8 in $g'$, 11 in $r'$ and 14 in $i'$). The variability is detected only in $i'$ filter, at the significance level of 0.001.

\begin{figure*}
\centering
\includegraphics[width=17cm]{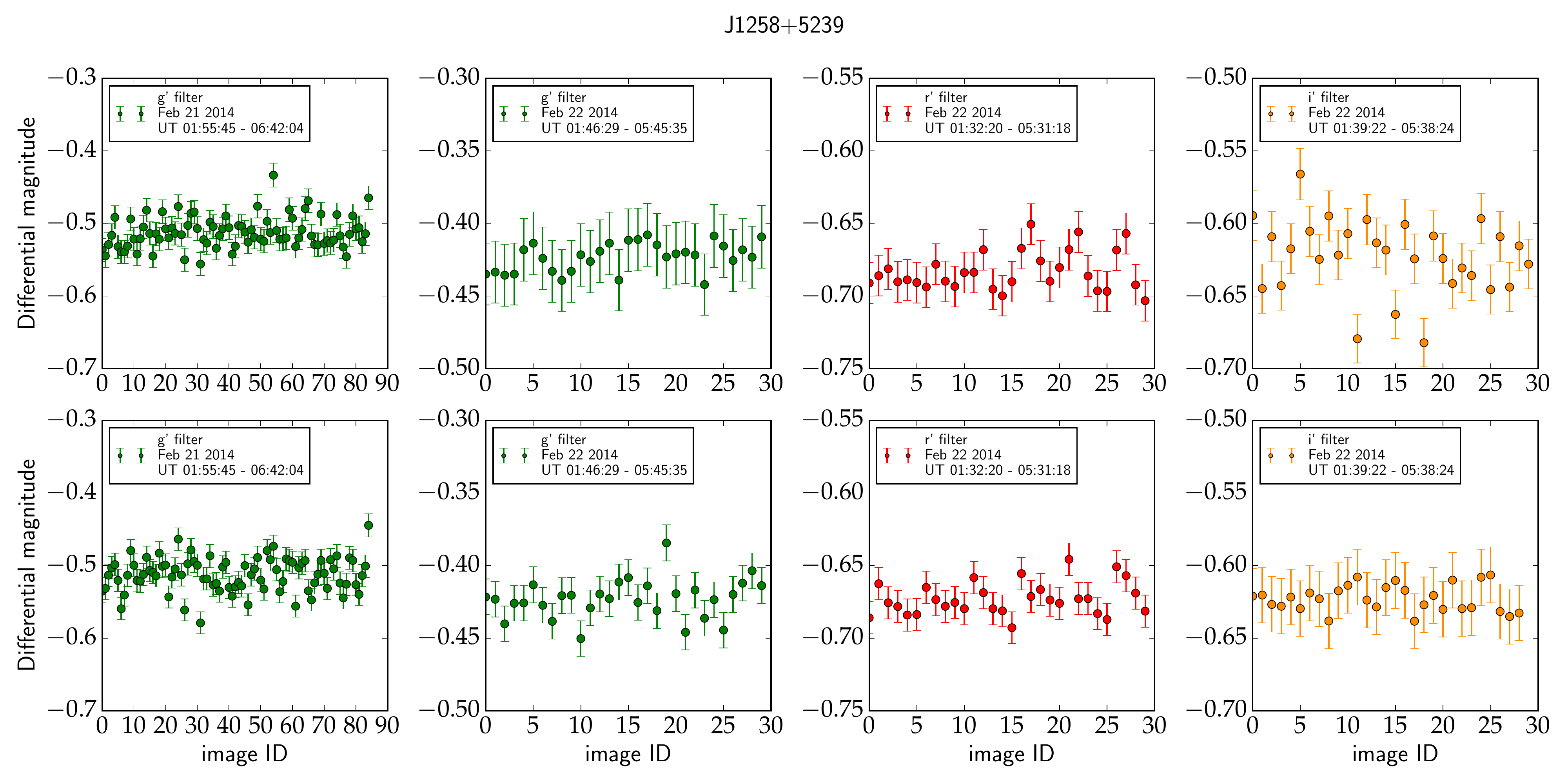}
\caption{Upper panels: Differential light curve for J1258+5239 during both nights in all available filters; lower panels: the same, transformed, for the non-variable check star used to control the test result. Variations were reported for February 22 in $i'$ filter with 0.001 significance level. \label{Fig6}.}
\end{figure*}

Considering that the variability was detected only during one night in one filter, we conclude that there is not enough evidence to confirm variable behavior in the source. 
Light curves for both observing nights of J1258+5239  are shown in figure \ref{Fig6}. As the microvariability was detected only in one filter, we conclude that we have not found enough evidence for microvariations during our observations of this source.

Given the presence of the Moon during the observations, the sensitivity of the observations might have been affected.

\textbf{J1316+4452} was observed during the second half of the nights, with the same conditions of lunar illumination as J1258+5239. Aside from our targeted source, there is a galaxy present in the field of view, southeast from the target as seen in Fig. \ref{Fig1}). It is of roughly similar brightness as our target. It is classified as a broad line galaxy in SDSS-III \citep{eis}. We have applied the same statistical tests on the galaxy as on the target. 

During the first observing night, we have obtained 84 images of the field (hence 84 observations of both, targeted type 2 quasar and accompanying galaxy) in $g'$ filter. We used 9 stars in the field for scaling. The total number of stars we have used for scaling was 580. Aperture photometry was measured within a diameter of 22.8 arcsec (60 px). The enhanced F-test on both, accompanying galaxy (J1316+4452b) and J1316+4452 itself results in negative detection of microvariability. 

{We have followed the same strategy as for the other targets and during the second observing night, we were circulating $g'$, $r'$ and $i'$ filters. We have obtained 30 exposures of the target in each filter. Usually there are not many comparison stars accompanying the quasar and the galaxy. However in this case, we were able to use 9 stars in $g'$, 8 in $r'$ and 9 in $i'$ filters. As the seeing improved compared to the first night, we have used aperture of diameter 20.8 arcsec (55 px). The statistical analysis result in detection of variability at the significance level 0.001 in both, broad-line galaxy and the type 2 quasar. The broad line sources (in this case quasars), were previously detected as variable sources on small scales \citep[e. g.][]{fic, pal}. Both light curves are displayed in figures \ref{Fig7} and \ref{Fig8} (respectively).

As in the case of J1258+5239, the observations were carried out in the presence of the Moon. As the only filter where we fail to detect variability is $g'$, which seem to be more  sensitive to the Moon shine, it is possible that the sources are indeed variable and the variability is not prominent enough to be detectable in $g'$ filter. Therefore we conclude that the source is probably variable.


\begin{figure*}
\centering
\includegraphics[width=17cm]{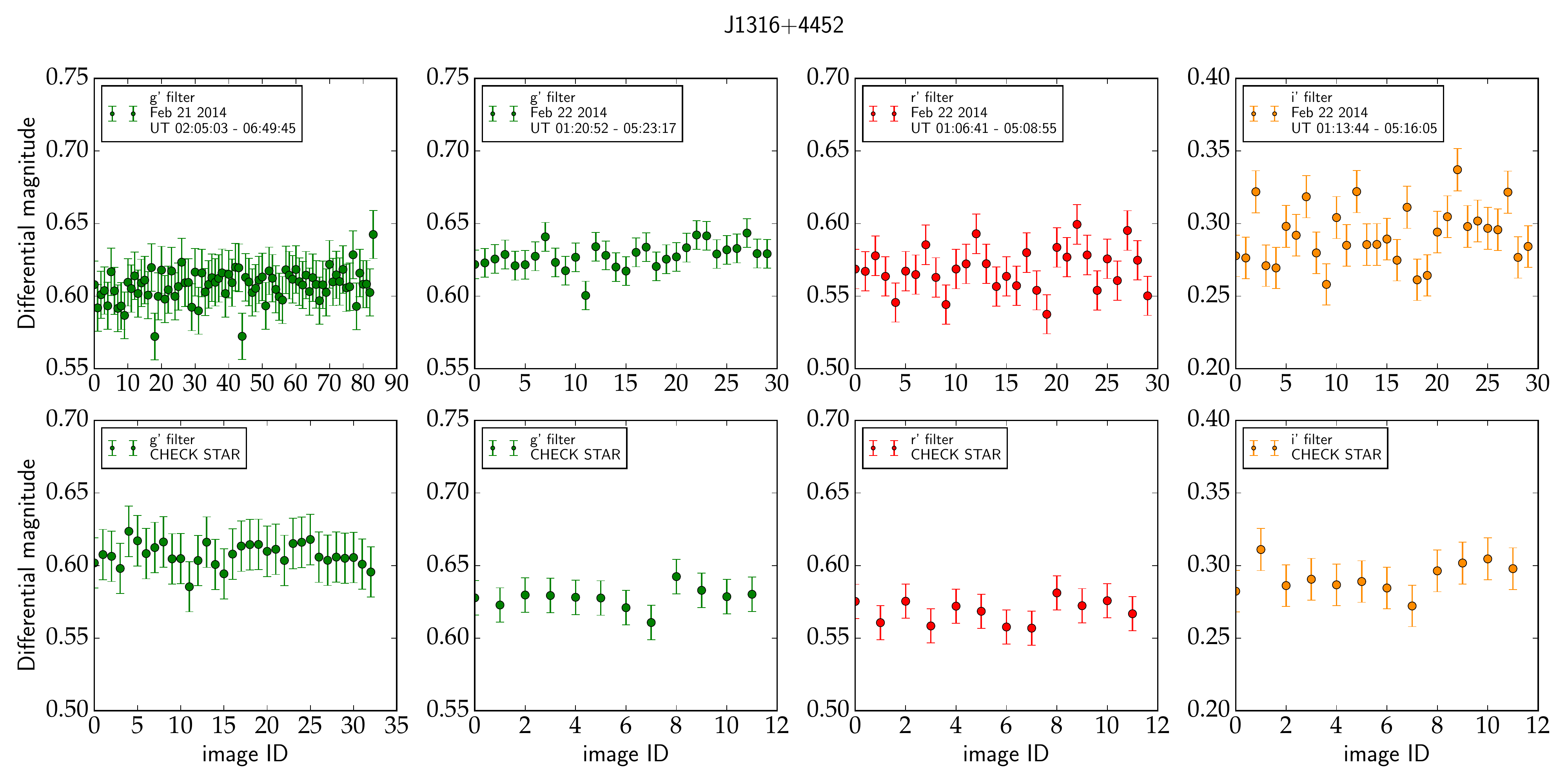}
\caption{Upper panel: differetial light curve for J1316+4452 during both nights in all available filters. We have detected microvariations at the 0.001 significance level in the $r'$ and $i'$ filters. Lower panels: transformed differential light curve of the check star. There are fewer observations of the check star, which is the result of the dithering pattern in combination with few stars in the field \label{Fig7}.}
\end{figure*}

\begin{figure}
\resizebox{\hsize}{!}{\includegraphics[width=17cm]{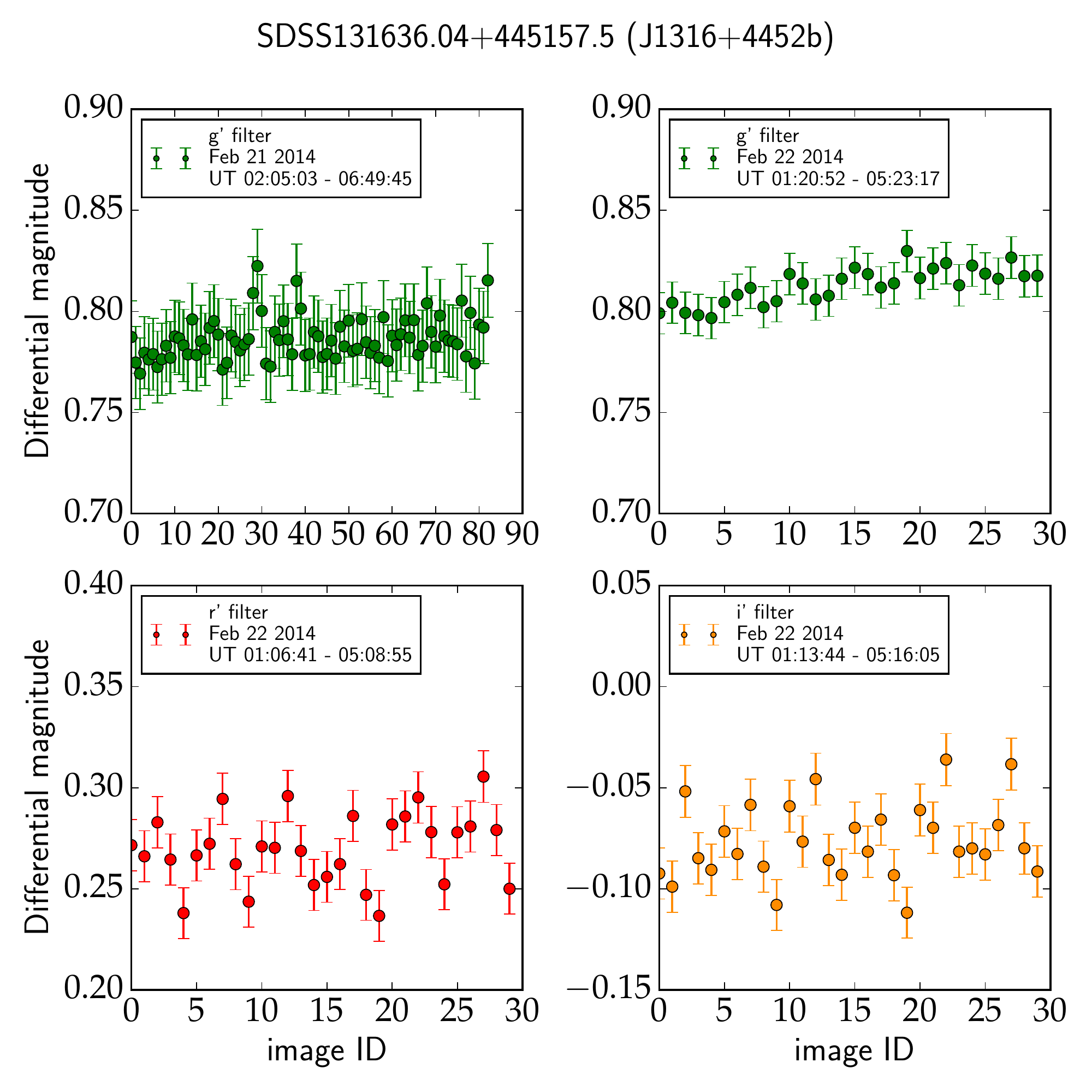}}
\caption{Differential light curve for the broad line galaxy J1316+4452b. The galaxy was observed in all the images obtained for the target, regardless of the dithering. It was detected as variable in $r'$ and $i'$ filter. There is no differential light curve for the check star plotted as it would be similar as in the preceeding figure \label{Fig8}.}

\end{figure}

\section{Discussion}

We have observed four type 2 quasars during two nights in February 2014 with the Nordic Optical Telescope located at the Roque de Los Muchachos observatory on La Palma, Canary islands. Out of four targets, two have disturbed morphologies. We have detected microvariability on a time scale less than 4 hours in one target (J0802+2552) in all filters at the 0.001 significance level during both observing nights, in J1258+5239 (in $i'$ filter) and J1316+4452 (in $r'$ and $i'$ filter) with 0.001 significance level only in the second observing night. In the cases of J1258+5239 and J1316+4452, the Moon was present during the observations, which probably resulted in lessening the sensitivity of the $g'$ and possibly $r'$ filter. As the phenomenon we are detecting is very fine, the smallest contribution from the Moon are likely affecting the observations.

J1258+5939 was detected as variable only in one filter and therefore we conclude there is not enough evidence for microvariability, J1316+4452  was detected as variable with a level of significance 0.01 in two filters, the negative detection is consistently in $g'$ filter which is affected by the Moon more than the $i'$ and $r'$. Therefore we conlude the target is probably variable. We are not drawing conclusions on one of the targets (J0843+3549) since the analysis was impossible due to the complex morphology of its host galaxy.

\subsection{F-test}

We have tested a newly proposed method for detecting variability/microvariability: the enhanced F-test \citep{die14} which was previously used only on simulated data and for detection of microvariability in blazars \citep{gaur2015}. The test shows sensible results and provides enough power to search for variability. But as the power of the test relies on the sample size, it is desirable to have as many observations as possible, and a high number of stars in the field observed. 

However there are limitations posed by the nature of the targets we have observed. As long as the target is only moderately extended without irregularities, the aperture photometry works reasonably well. Irregular targets are problematic as, at the moment, there is no routine available which would be able to separate the quasar from its host galaxy. Including the presumably steady host galaxy, increases the noise in the data, which results in a possible underestimation of the amplitude of the variation. We have shown that in the case of significantly extended host galaxies (as in some interacting systems), the variability detection is hampered due to the problems with aperture photometry.
Alternative photometry available in SExtractor (e.g. isocor photometry or elliptical photometries using Kron radius) might introduce inhomogeneities into the dataset, due to different aperture sizes for the stars in the field and the quasars, therefore we are not focusing on them. We have tried to limit the threshold to account for light propagating into the main target from the fainter tail of the source. Such adjustment severely reduces the number of stars that can be used for differential photometry and, therefore, diminishes the main advantage of the F-test, which is to increase the number of degrees of freedom based on using as many field stars as possible. 
We have shown that the brightness of the sky might play a role in the reliability of the test as we were unable to achieve similar photometric precisions for targets observed when the sky was partially illuminated by the Moon.

\subsection{Microvariability results}

Optical microvariability in type 2 quasars is a phenomenon previously unaccounted for. It is hard to observe short time scale variability in this kind of objects since our view to the central engine is supposed to be blocked by the dusty obscuring torus. 

Possible explanation of this phenomenon might lie in the misclassification of the targets, caused by the insufficient quality of the spectra used for analysis. Such misclassification can be addressed by infrared data as studied by \citet{ches}. They suggested that the population of type 2 quasars might be divided into a group of true type 2 quasars without any underlying broad line regions and heavily obscured type 1 quasars, based on the infrared color-color diagram. The study by \citet{ches} showed that 43\% of type 2 quasars are dominated by a thermal spectrum, while the rest is following a power law, pointing out a non-thermal origin of the emission, which bears a similarity with type 1 quasars. Out of our studied sample, one target (J13116+4452) falls into the power law group. This target is detected as probably variable which is consistent with the results of microvariability detections in type 1 sources \cite[e. g.][]{fic, pal}. Such result hints that the belonging to the power law group, based on the infrared color, might have an effect on microvariability detection. Drawing conclusion based on one source is however not reliable and we would need a larger sample to confirm the hypothesis.

The only target in which we have detected microvariability during both nights in all filters stands out in the sample at first sight because of its morphology. Since all our targets are selected based on the SDSS catalog, we were able to check the morphology types based on citizen science projects Galaxy ZOO and Galaxy ZOO2 \citep{lin, wil}. All of our targets have been classified by more than 30 Galaxy ZOO users. In the first iteration, J1258+5239 is the only one which reaches the reliability threshold for the classification. It is classified as spiral galaxy, while the other two (J0802+2552 and J1316+4452) are uncertain. The second iteration of the morphology classification marked J1316+4452 as a spiral with a bulge and medium winded arms with notable irregularity, while J0802+2552 was marked as an irregular elliptical with many votes in favor of a merging state. However since our sample is very small, it would be premature to draw conclusions on a relation between the detection of microvariability and the morphology.

The physical processes lying behind the microvariations are difficult to determine as there are not many reports of similar phenomena to this date. Recent studies dedicated to rapid variations are directed mostly on local AGNs in X-ray \citep{ris05, ris09}. \citet{ris09} observe changes on the time scales of hours in NGC 1365, a Seyfert 1.8 galaxy . They report a transition from the reflection dominated source to increase in the 7-10 keV emission in $\sim$ 10 hours. The authors propose that the variations are due to the rapid changes in the absorbing column density along the line of sight. Although we cannot confirm such behavior in our target, as we have no X-ray data available, we cannot rule out the possibility of a similar behavior in our case. Detected microvariations in the optical regime would fall into the gaps where there is no Compton-thick clouds obscuration. We can assume that such change would be reflected in the optical spectra, although to the best of our knowledge, there was no rapid optical spectroscopic follow-up performed on the NGC 1365. Changes in the spectrum was not reported in our microvariable target either.
The explanation based on the window in the obscuring medium is consistent with the clumpy model of the torus as proposed by e. g. \citet{nen}, which includes clouds of obscuring material passing through the line of sight, occasionally allowing the view onto the central engine. As microvariability is common in the unobscured targets, the behavior of our targets seem to be consistent with such scenario.

\section*{Acknowledgements}

Based on observations made with the Nordic Optical Telescope operated on the island of La Palma
by the Nordic Optical Telescope Scientific Association in the Spanish Observatorio del Roque de Los Muchachos of the 
Instituto de Astrof\'isica de Canarias. 
This research has been supported by the Spanish Ministerio de Econom\'ia y Competitividad (MINECO) under the grant AYA2014-58861-C3-1.
IO acknowledges support from the European Research Council (ERC) in the form of Advanced Grant, {\sc cosmicism}.
JAD is grateful for the support from the grant UNAM-DGAPA-PAPIIT IN110013 Program and the Canary islands CIE: Tricontinental Atlantic Campus. 
AE acknowledges support by the grant AYA2012-30789.




\bibliographystyle{mnras}
\bibliography{bibliography.bib} 








\bsp	
\label{lastpage}
\end{document}